\documentclass[prl,twocolumn,aps,showpacs,amsmath,amssymb]{revtex4-1}

\usepackage{graphicx}
\usepackage{dcolumn}
\usepackage{bm}

\begin{document}

\title{Reversible modulation and ultrafast dynamics of THz resonances \\
in strongly photoexcited metamaterials}

\author{I. Chatzakis,$^{1}$ L. Luo,$^{1}$ J.
Wang,$^{1,\dagger}$ N-H. Shen,$^{1}$ T. Koschny,$^{1}$
J. Zhou,$^{2,\ddagger}$}\author{C. M. Soukoulis$^{1,3}$}

\affiliation{$^1$Ames Laboratory and
Department of Physics and Astronomy, Iowa State University, Ames,
Iowa 50011, U.S.A.\\ $^2$Center for integrated Nanotechnologies, Los Alamos National Laboaratory, Los Alamos, New Mexico 87545, USA.\\
 $^3$Institute of Electronic Structure and Laser, FORTH,
71110 Heraklion, Crete, Greece}

\date{\today}

\begin{abstract}
We demonstrate an ultrafast reversible modulation of resonant
terahertz (THz) response in \emph{strongly photoexcited}
metamaterials. The transient spectral-temporal response of the
dipole transition $\sim$1.6 THz exhibits a distinct non-monotonic
variation as a function of pump fluence. The transition energy
shift, strength, spectral width and density-dependent ultrafast
relaxation manifest a remarkable re-emergence of the resonances
after initial quenching. Our simulation, incorporating the
first-order diffraction from the photoinduced transient grating,
reproduces the salient features, providing a new avenue for
designing nonlinear and frequency-agile THz modulators.

\end{abstract}

\pacs{78.67.Pt, 78.20.-e, 78.47.-p, 42.25.Bs}

\maketitle

Artificially subwavelength structured materials, so-called
metamaterials
\cite{Smith2004,Soukoulis2007,Shalaev2007,Soukoulis2011}, attract
strong current interest due to their exceptional properties, such as
simultaneously negative permittivity $\varepsilon $($\omega $) and
permeability $\mu $($\omega $), which implies a negative refractive
index
\cite{Smith2004,Soukoulis2007,Shalaev2007,Soukoulis2011,Shelby2001}
not available in natural materials. In addition, artificial
magnetism \cite{Yen2004,Linden2004}, super focusing
\cite{Fang2005,Pendry2000} and specifically-tailored structures
responding to ultra-broadband electromagnetic radiation, from
gigahertz (GHz), terahertz (THz) to visible range
\cite{Soukoulis2011,Zhang2005}, set them among the most promising
candidates for next generation optoelectronic devices and
large-scale photonic functional systems.

Recently, there has been significant interest in understanding
ultrafast THz responses of nonlinear metamaterials, e.g., split ring
resonators (SRRs) patterned on substrates exhibiting substantial
nonlinearities \cite{Azad2006}. The resulting components and schemes
allow for constructing controllable active THz devices with
dynamical tunability of amplitude and phase, which fills the
demanding THz technology gap. Prior experiments using time-domain
THz spectroscopy have demonstrated dynamic tuning of both magnetic
\cite{Chen2006,Padilla2006,Shen2011,Chen2007,Chen2008,Chen2011} and
electric \cite{Manceau2010} dipole resonances of SRRs on semiconductors or
superconductors. More complex structures of coupled resonators on
GaAs substrates have been proposed and experimentally studied,
revealing frequency-agile THz modulation, e.g., a resonance shift
with increasing photoexcitation \cite{Chen2008,Shen2011}. However,
thus far all ultrafast studies of THz materials have been
concentrated on the relatively low photoexcitation, where a
\emph{reduction} or \emph{quenching} of the resonant absorptions is
observed \cite{Shen2011,Padilla2006,keshav2009}. Two outstanding
issues are still poorly understood: first, there lack insights on
how \emph{high density} photoexcited carriers modify the dynamic
responses of metamaterials; second, the decay pathway of
photoexcited metamaterials after femtosecond (fs) excitation is not
explicitly studied, and, particularly, the temporal evolution of the
transient states have only been understood as a
density-\emph{independent} relaxation back to the equilibrium.

In this letter, we demonstrate an ultrafast reversible modulation of
resonant THz absorption in \emph{strongly photoexcited}
metamaterials by optical pump and THz probe spectroscopy. This
process directly manifests itself via a remarkable reemergence of
the originally quenched THz resonance above a crossover density
$N_{c}$$\sim$$4.7\times 10^{16} {\rm \; cm}^{-3} $ after femtosecond photo-excitation. Increasing the excitation from below to above the
transition density, we identify two distinctly different relaxation
pathways of the THz resonance with opposite dynamics. A model
calculation of transient spectra incorporating the first order
diffraction mode from the photoinduced transient grating reproduces
the salient features, which is further corroborated by their
dependence on the unit cell lattice constant. The revealed scheme
represents a relatively simple and generic approach to achieve nonlinear and
frequency-agile functions in THz device without particularly complex
structures.

The samples used are double SRRs on GaAs, illustrated in the inset
of Fig. 1 (a), which are patterned 6 $\mu$m copper rings on high
resistivity GaAs of 630 $\mu$m thick. We mainly focus on two SRR
samples with unit cell lattice constant of 50 $\mu$m and 45 $\mu$m,
and the outer dimension of an individual SRR is 36 $\mu$m. Our
optical pump and THz probe spectroscopy setup is driven by a 1 kHz
Ti:sapphire regenerative amplifier with 40 fs pulse duration at
800 nm. One part of the output is used to excite the sample, while
a small fraction is used to generate and detect THz pulses \cite{Wu1996}.
 Phase-locked THz field transients are used as a probe, which is generated
via optical rectification and detected by electro-optic sampling
in 1 mm ZnTe crystals (supplementary material). 
Copper apertures of 2.5 mm are placed in front of the sample and a
GaAs reference to ensure a uniform illumination and faithful comparison. The THz field
transmission coefficient $T(\mathop{\omega }$) is obtained by the
ratio between two Fourier transformed spectra from the transmitted
THz probe pulses through the SRRs and bare GaAs. The transient THz
signals are recorded at various time delays with respect to the pump
pulse to obtain time-dependent $T(\mathop{\omega }$). The whole
setup is purged with dry-N${}_{2}$ gas.

\begin{figure}[tbp]
\begin{center}
\includegraphics[scale=0.36]{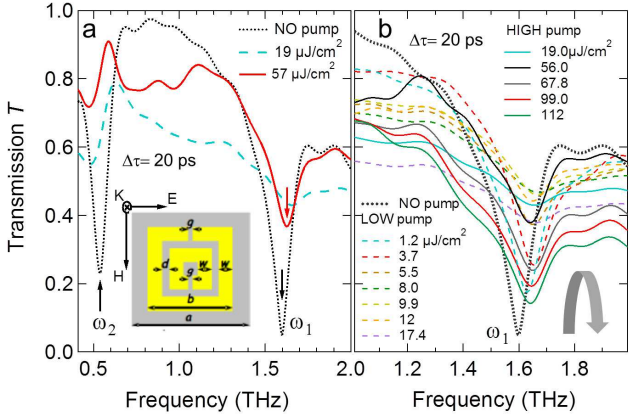}
\caption{(color online)(a): Transient transmission spectra
$T(\omega)$ taken under two excitation fluences, 19 $\mu
$J/cm${}^{2}$ (cyan dashed) and 57 $\mu $J/cm${}^{2}$ (red solid),
respectively. Static transmission spectra without pump is plot
together (black dotted). The inset: the SRR unit cell of our sample
and the polarization of the normally incident THz radiation. a$=$50
$\mu$m, b$=$36 $\mu$m, d$=$3 $\mu$m, g$=$2 $\mu$m and w$=$6 $\mu$m.
(b): The non-monotonic  pump fluence dependence (indicated by the
gray arrow) of the resonant absorption $\sim$1.63 THz for low
(dashed lines) and high (solid lines) photoexcitation. Spectra In
both panels are taken at fixed time delay of 20 ps. } \label{Fig1}
\end{center}
\end{figure}

The static transmission spectra are shown in Fig. 1(a) (black dotted)
with a THz probe normally incident on the sample plane. The \emph{E}
field of the probe is parallel to the gaps of SRRs (inset). There are two
main THz resonances detected in this configuration at
$\omega_{1}\sim$1.6 THz and $\omega$${}_{2}$$\sim$ 0.5 THz, which
originate from resonant electric processes consistent with the prior
measurements \cite{Padilla2006,Manceau2010}. The $\omega_{1}$ can be
understood as the electric dipole resonance of the metallic bars
of the SRRs, and the $\omega_{2}$ is the magnetic dipole resonance induced by the circulating
electric currents generated by the incident \emph{E} field.

The two THz resonances exhibit distinctly different pump fluence
dependence, as shown in the transient THz transmission spectra at
$\Delta\tau=$20 ps (Fig. 1(a)) for two pump fluences: 19 $\mu
$J/cm${}^{2}$ (cyan dashed) and 56 $\mu $J/cm${}^{2}$ (red solid).
The 20 ps time-delay between the excitation and probe pulse is
introduced to ensure a measure of quasi-steady state considering the
transient carrier lifetime in GaAs on the order of 1 ns. The 19 $\mu
$J/cm${}^{2}$ trace exhibits largely quenched resonances for both
the $\omega_{1}$ and $\omega_{2}$ after the photoexcitation.
However, the 56 $\mu $J/cm${}^{2}$ trace clearly shows substantial
recovery of the resonant THz absorption around $\omega_{1}$, while
the $\omega_{2}$ resonance remains quenched. This salient feature is
further corroborated in the detailed pump fluence dependence shown
in Fig. 1(b). In the low excitation regime from 1.2 to 19
$\mu$J/cm${}^{2}$, there is a clear weakening in the transition
strength and broadening in the linewidth for the $\omega_{1}$
resonance, with an almost complete quenching at pump fluence of 19
$\mu$J/cm${}^{2}$. Here the transmission is $\sim$45\% higher and
the peak energy has a blue shift of $\sim$60 GHz. In strong contrast
to this, futher increasing photoexcitation from 56 $\mu$J/cm${}^{2}$
to 112 $\mu$J/cm${}^{2}$ results in a progressively pronounced
resonance. The spectra clearly show a red shift of the resonance energy, $\sim$20 GHz for the highest pump fluence used, a partial
recovery of the transition strength and linewidth narrowing. This
remarkable non-monotonic behavior and re-emerging resonance near the
$\omega_{1}$ manifest a reversible and frequency-agile modulation of
the resonant THz absorption by photoexcitation. This behavior has
been seen experimentally for the first time, and we can explain below
that for very high pump fluence the first-order diffraction will surpass the electric resonance behavior in the THz responses of photoexcited metamaterials. 
Fig. 1(b), the dips in the transmission above 56 $\mu
$J/cm${}^{2}$ are due to the first-order diffraction peaks, which will be discussed further below.

\begin{figure}[htb]
\includegraphics[width=8.5 cm]{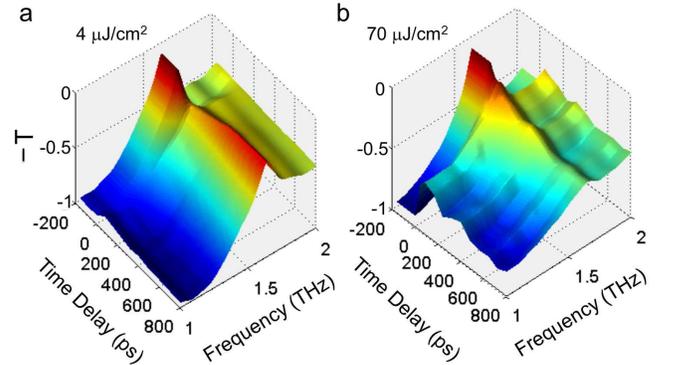}
\caption{(color online) Time evolution of negative transmission spectra, -T($\omega$), in the photoexcited metamaterial of 50$\mu$m lattice constant for
(a) 4 $\mu $J/cm${}^{2}$ and (b) 70 $\mu $J/cm${}^{2}$,
respectively. These clearly show the opposite relaxation pathways
for the weak and strong excitation regimes (see text).} \label{Fig1}
\end{figure}

The weak and strong photoexcitations lead to significantly different
relaxation pathways back to the equilibrium, as shown in the time
evolution of T($\omega$) spectra following photoexcitation in Figs. 2(a) and
2(b). For pump fluence at 4 $\mu$J/cm${}^{2}$, the THz absorption
$\sim$$\omega_{1}$ resonance exhibit an initial photoinduced
reduction, immediately followed by a relaxation back to the original
vaule on a ns time scale. Interestingly enough, opposite
to those at the low excitation, the 70 $\mu$J/cm${}^{2}$ dynamics
clearly shows a further reduction of the THz resonence $\sim$1.6 THz
following the initial decrease. We emphasize two key aspects of this
observation: (i) at the large time delay of $\Delta\tau=$800 ps, the
THz resonance $\sim$1.6 THz is almost completely quenched for the 70
$\mu$J/cm${}^{2}$ excitation, although a clear recovery of the
resonance has been seen for the 4 $\mu$J/cm${}^{2}$ case; (ii) the
pump-fluence-dependent relaxation dynamics of the resonance
indicates another way to achieve nonlinear and reversible THz
modulation by tuning the time delay.

\begin{figure}[tbp]
\includegraphics[scale=0.4]{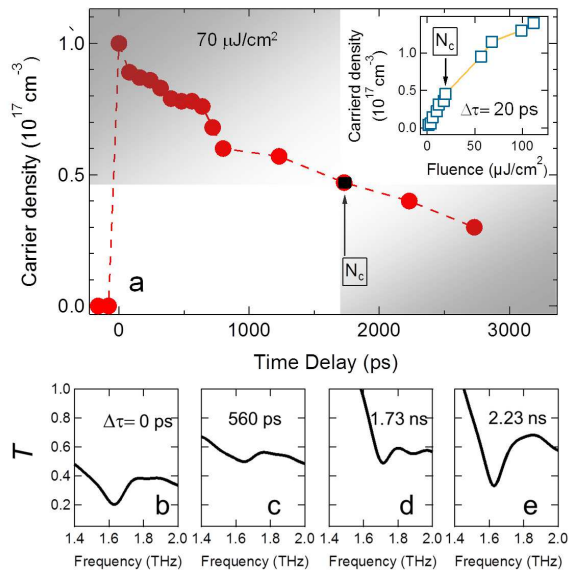}
\caption{(color online)(a) The extracted transient carrier
concentration as function of time delay. The left and right gray
areas illustrate a crossover from the high to low excitation regime
with a crossover density N$_{c}$ (black square). The inset plots the
extracted transient carrier concentration as function of pump
fluence at $\Delta\tau=$20 ps, marked together with N$_{c}$
corresponding to the density excited by 19 $\mu $J/cm${}^{2}$.
(b)-(e): The transient transmission spectra T($\omega$) at various
time delays following 70 $\mu $J/cm${}^{2}$ photoexcitation.}
\label{Fig1}
\end{figure}

Ultrafast photoexcitation strongly alters the electronic states in
GaAs substrate by injecting non-equilibrium transient carriers
during the pulse duration of 40 fs. Subsequent
carrier-carrier/carrier-phonon collisions lead to decoherence and
then to a quasi-thermal, hot transient carrier distribution within 1
ps. This hot carrier distribution further cools down by emitting
phonons and eventually relaxes back to the equilibrium by carrier
recombination on the order of 1 ns. In the weak excitation regime below
19 $\mu $J/cm${}^{2}$ (see Fig. 1), the injected carriers make the GaAs regions  in the gaps of SRRs and in between them conductive. This leads to the reduction of the resonant absorption near $\omega_{1}$  and even complete quenching of the resonances \cite{Padilla2006}. Thus the observed crossover from the
photoinduced quenching to re-emergence of the THz resonance clearly
indicate a new collective excitation setting in at the high
excitation regime, not accessible in the previous measurements.

To gain quantitatively insights into the transition density $N_{c}$
for the excitation-induced crossover behavior, we experimentally
determine temporal evolution of the transient carrier density. 
 The phase-sensitive nature of the time-domain THz field measurement can directly yield both real and imaginary parts of the optical conductivity in the photoexcited GaAs substrate, on equal footing without any model assumptions. 
This gives a direct determination of
the transient carrier density based on an analysis using the Drude
model (see supplementary material). 
Figure 3(a) plots the temporal
evolution of the transient carrier concentration excited by 70 $\mu
$J/cm${}^{2}$, which shows an exponential decay with a peak density
10${}^{17}$ cm${}^{-3}$.  By correlating the temporal profile to the
fluence-dependent carrier concentration at $\Delta\tau=$20 ps (inset,
Fig. 3(a)), one can define a crossover point at density
N$_{c}$$\sim$$4.7\times 10^{16} {\rm \; cm}^{-3} $ at time delay of
$\Delta\tau_{c}$$\sim$1.73 ns (black square), corresponding to the
transient density for 19 $\mu $J/cm${}^{2}$. This
naturally divides the transient electronic response of photoexcited
metamaterials into two regimes with opposite time and excitation
dependence, as illustrated in the shaded areas of Fig. 3(a).
Transient spectra at various time delays, as shown in Figs.
3(b)-(e), show a reversal of the THz transmission change once the
system is driven cross the boundary, corroborating well of our
conclusion, e.g., the THz transmission recovers its strength at
$\Delta\tau=$2.2 ns after initial quenching at time delays before
1.73 ns.

Next, we analyze the transient THz response of photoexcited
metamaterials by full-wave numerical simulations. Before we take
into account any complex processes, we first simply consider the
photoexcitation-induced dynamic response of the GaAs substrate
beneath the SRRs. 
A photoexcited layer is modeled by the well-established Drude model, in which, the
frequency-dependent complex conductivity $\sigma ={\varepsilon _{0}
\omega _{p}^{2} \mathord{\left/ {\vphantom {\varepsilon _{0} \omega_{p}^{2} (\gamma -i\omega )}} \right. \kern-\nulldelimiterspace}(\gamma -i\omega )} $.
Here, $\varepsilon _{0} $ is the permittivity
of vacuum, $\gamma $ represents the collision frequency, and $\omega
_{p} =\sqrt{{Ne^{2} \mathord{\left/ {\vphantom {Ne^{2}  (\varepsilon
_{0} m^{*} )}} \right. \kern-\nulldelimiterspace} (\varepsilon _{0}
m^{*} )} } $ the plasma frequency. Some other parameters include the
carrier density \textit{N}, the free electron charge \textit{e}
($1.6\times 10^{-19} $C) and the effective carrier mass in GaAs,
$m^{*} =0.067m_{0} $ (\textit{m}${}_{0}$, the mass of a free
electron). Thus, the photoexcited layer has the dielectric function
$\varepsilon (\omega )=\varepsilon _{s} +{i\sigma \mathord{\left/
{\vphantom {i\sigma (\varepsilon _{0} \omega )}} \right.
\kern-\nulldelimiterspace} (\varepsilon _{0} \omega )} $, where
$\varepsilon _{s} $ (=12.7) is the dielectric constant of undoped
GaAs. In our simulations, we keep $\gamma =1.8$ THz for all the
cases as in Ref. 17.

Figure 4(a) shows the simulated transmission spectra for various
carrier density \textit{N}. The simulation clearly shows the
crossover from a reduction to an enhancement of the resonant THz
absorption $\sim$1.6 THz as the excitation from low to high density
regime. This agrees well with the photoinduced reversible modulation
demonstrated in the experiments. Specifically, in the low density
below $4.0\times 10^{16} {\rm \; cm}^{-3} $, the THz resonance peak
in the transmission curve exhibit clearly weakened transition
strength, a blue-shift and spectral broadening. Increasing the
carrier density above $4.0\times 10^{16} {\rm \; cm}^{-3} $, the
resonant THz peak progressively reemerges, e.g., a clear
transmission dip seen for the case $N=8.1\times 10^{17} {\rm \;
cm}^{-3} $. From the simulations, we can attribute the reemergence
of the resonant THz absorption at high density to the effect of
first-order diffractive waves from photoinduced transient grating.
Due to the periodic structure of the SRRs, the first-order
diffractive wave is found to propagate in the substrate at high
excitation, which strongly influence the fundamental absorption
mode. 
This leads to a new THz transmission dip at $f_{1} =c/(na)$
above a transition density $\sim$ $4.0\times 10^{16} {\rm \;
cm}^{-3} $. Here \textit{c} is light speed in vacuum, \textit{n} the
refractive index of GaAs and \textit{a} the lattice constant of the
structure. The SRR structure with the dipole resonance close to
$f_1$ allows to achieve a reversal of resonant THz absorption with
increasing photoexcitation, due to a crossover in the absorption
from the dipole mode to the diffraction mode.

To corroborate our model, we compare the re-emerged THz resonances
from the experiment (Fig. 4(b)) to those from the simulation (Fig.
4(c)) for two different lattice constants, 50 $\mu$m and 45 $\mu$m,
respectively. The simulations are performed to the structures with
the same double-SRR array for the case with the density $N=8.1\times
10^{17} {\rm \; cm}^{-3} $.  It is clearly visible that the
simulated transmission dip to higher frequency by decreasing the
lattice constant, and the dip positions tightly link to the
frequency of the first-order diffraction mode for each case. An
extremely good agreement between the simulation and the experiment
is found for the transmission-dip position and its shift, which
underpins the effect of the high-order diffraction mode in the
strongly excited metamaterials.
\begin{figure}[tbp]

\includegraphics[scale=0.38]{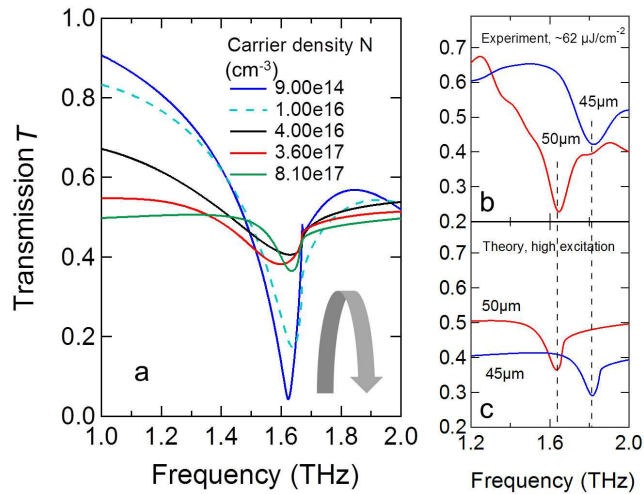}
\caption{(color online) (a): The simulated THz transmission spectra
for various carrier densities \textit{N}. The photoinduced THz
resonances in the strong excitation regime for two different lattice
constants 45 and 50 $\mu$m: (b) experiment and (c) simulation. }
\label{Fig1}
\end{figure}

In summary, we have demonstrated ultrafast photoinduced reversible
modulation of resonant THz absorption in strongly photoexcited
metamaterials. Increasing the excitation from below to above the
threshold density N$_{c}$, we observe a crossover from a complete
quenching to a reemergence of the THz resonance. Our analysis and
theoretical simulations, based on the first order propagation from
the photoinduced transient grating, explain the density- and
lattice-constant-dependent frequency shift.  Our results clearly
identify the importance of photoinduced optical modes in
metamaterials, besides the electronic resonances, which should be
carefully considered in designing future multifunctioning
photonic-electronic devices using the artificial periodical
structures.  The easy implementation of the revealed scheme
represents another approach to achieve nonlinear and frequency-agile
functions in THz devices.

Work at Ames Laboratory was supported by the Department of Energy
(Basic Energy Sciences) under contract No. DE-AC02-07CH11358. This
was partially supported by IC Postdoctoral Fellowship Program.

\medskip

\noindent$^{\dagger}$To whom correspondence should be addressed.
Electronic address: jgwang@iastate.edu

\noindent$^{\ddagger}$Present address: Department of Physics, University of South Florida, Tampa, FL

\bibliographystyle{apsrev}

\newpage

\end{document}